\documentstyle[preprint,aps]{revtex}
\newcommand{\Nu}{{\cal N}}
\newcommand{\Nur}{{\cal N}_r}
\newcommand{\Nui}{{\cal N}_i}
\newcommand{\Dv}{{\bf D}}
\newcommand{\Ev}{{\bf E}}
\newcommand{\Hv}{{\bf H}}
\newcommand{\jv}{{\bf j}}
\newcommand{\nv}{{\bf n}}
\newcommand{\qv}{{\bf q}}
\newcommand{\vv}{{\bf v}}
\newcommand{\nabv}{{\mbox{\boldmath{$\nabla$}}}}
\newcommand{\intd}{\int\!\!\!\int}
\newcommand{\eqref}[1]{eq. (\ref{#1})}
\newcommand{\xunit}{{\bf {\hat x}}}

\newcommand{\zunit}{{\bf {\hat z}}}
\begin{document}
\draft
\title{Electric Nusselt number characterization of electroconvection in
nematic liquid crystals}
\author{J.T. Gleeson, N. Gheorghiu}
\address{Department of Physics, \\
Kent State University\\
Kent, OH 44242 USA}
\author{E. Plaut}
\address{Laboratoire d'Energ\'{e}tique et de M\'{e}canique Th\'{e}orique et 
Appliqu\'{e}e, \\
Institut National Polytechnique de Lorraine\\
54504 Vandoeuvre cedex, France}
\date{\today}
\maketitle
\begin{abstract}
We develop a characterization method of
electroconvection structures in a
planar nematic liquid crystal layer
by a study of the electric current transport.
Because the applied potential difference has
a sinusoidal time dependence, we
define two electric Nusselt numbers corresponding
to the in-phase and out-of-phase components of the current.
These Nusselt numbers are predicted theoretically
using a weakly nonlinear analysis of the standard model.
Our measurements of  the electric current
confirm that both numbers vary linearly with the distance from onset 
until the occurence of secondary instabilities;
these instabilities also have a distinct
Nusselt number signature.
A systematic comparison between our theoretical and experimental results,
using no adjusted parameters, demonstrates reasonable agreement.
This represents a quantitative test of the standard
model completely independent from traditional, optical techniques
of studying electroconvection. 
\end{abstract}
\narrowtext

Although spontaneous pattern formation within structureless
environments pervades nature
\cite{cross}, a comprehensive understanding of this complex behavior
remains elusive.
Therefore well-controlled experimental systems exhibiting pattern formation
are extensively studied; among these,
thermoconvection of a layer of fluid heated from below
\cite{benard-rayleigh}
and electroconvection of a nematic liquid crystal layer
\cite{standardmodel,kramer1}
are particularly interesting since they allow very
large aspect ratio geometries.
In both of these systems,
convection structures
form spontaneously when the 
applied stress,
i.e. the gradient of either the temperature
or the electric potential,
exceeds a critical value.
These inherently non-equilibrium structures can
only persist when there is an energy source to overcome the
dissipation associated with the flow.
Therefore,
energy transport studies represent a particularly valuable technique for
elucidating the essential nature of the instabilities that lead to
these patterns.
For example, the first accurate determination
of the stress necessary to induce
thermoconvection was made by measuring the heating power
required to sustain a desired temperature difference across a thin layer
of water
\cite{schmidt}.
This power is customarily expressed as the Nusselt number, defined as
the heat flow across a fluid layer relative to 
the heat flow required in the absence of fluid flow.
Nusselt number measurements remain a method
of choice for studies not only of the structured states
that occur during
thermoconvection when the
stress is only slightly above its critical value
\cite{ahlers1}
but also of the turbulent flow that occurs when the
stress is enormous
\cite{chavanne}.
By contrast the electroconvection of a planar nematic liquid crystal layer,
which represents a similar but fully \textit{anisotropic}
model pattern forming system,
has previously been studied only with qualitative or semi-quantitative
optical techniques.
Reports of energy flow measurement during electroconvection are rare
\cite{Rehberg1},
and no theoretical studies of the energy transport exist for this system.
The aim of this Letter is to fill this gap.

Electroconvection is obtained when an a.c. electric potential,
$\sqrt{2}V\cos(\omega t)$, is applied
to two horizontal ($\perp\zunit$) electrodes separated by distance $d$
confining a nematic liquid crystal.
Here we focus on the \textit{planar} anchoring
case where the director field $\nv$
is fixed to $\xunit$ at the confining electrode plates.
The instability relies on a
coupling between $\nv$, the velocity field, $\vv$
and the induced charge density, $\rho_e$ or equivalently the induced
electric potential
$\phi$, such that the full electric field reads
$\Ev={\sqrt{2}V}/{d}~[\cos(\omega t)\zunit-d\nabv\phi\big]$.
At moderate frequencies
$\omega$, when
$V$ exceeds a critical value $V_c$~,
the instability sets in the form of
normal \textit{conduction} rolls of wavevector $\qv=q\xunit$~;
at large frequencies \textit{dielectric} rolls are observed \cite{kramer1}
but we do not consider this regime in this work.
These phenomena are well explained via the standard model (SM)
for electroconvection
\cite{standardmodel,kramer1}
where the charge conduction in the liquid crystal is assumed Ohmic.
In addition to linear properties (values of $V_c$ and $q$
as a function of $\omega$),
the SM explains several secondary instabilities that are
experimentally observed, such as the transitions to zig-zag rolls,
stationary and oscillatory
bimodal patterns and abnormal rolls \cite{plaut-pesch-99}.
The one phenomenon which the
SM has been unable to predict is the
traveling roll state. For this, a new approach,
the weak electrolyte model, was developed,
in which the electrical conductivity is
assumed to be due to two species of dissociated ions
having different mobilities
\cite{treiber}.
With this model a semi-quantitative agreement with experimental results
on traveling rolls was demonstrated \cite{treiber}
but fitting parameters were necessary.
Because traveling rolls are not encountered at low frequencies
where we performed our experiments, and
because the SM is much less complicated, we did not use the 
weak electrolyte model.

The total current $I$ through the nematic cell
enclosed by the horizontal electrodes of area $S$ can be
calculated as the circulation of the magnetic induction
$\Hv$. From the
Maxwell-Amp\`ere equation,
$\nabv\times\Hv=\jv+\partial_t\Dv$,
$I$ is sum of the conduction and displacment currents:
\begin{equation}
 \label{current_gene}
 I~=~\intd_S \big(j_z\ \, + \, \partial_t D_z\big)\ dx\ dy
\end{equation}
where, within the SM,
$\jv ~=~\sigma_\perp\Ev+\sigma_a(\nv\cdot\Ev)\nv + \rho_e\vv$~;
$\Dv~=~\epsilon_\perp\Ev+\epsilon_a(\nv\cdot\Ev)\nv$~;
$\sigma_\perp$ ($\sigma_\parallel$) are the conductivities perpendicular
(parallel) to $\nv$~, and,
$\epsilon_\perp$ ($\epsilon_\parallel$)
are the dielectric
permittivities perpendicular (parallel) to $\nv$~,
$\sigma_a=\sigma_\parallel-\sigma_\perp$ and
$\epsilon_a=\epsilon_\parallel-\epsilon_\perp$ are the corresponding
anisotropies.
Note that the surface integral in \eqref{current_gene}
does not depend on the $z$-value ($-d/2\le z\le d/2$) chosen,
because of the Maxwell-Amp\`ere equation.
In the quiescent (no convection) state $\nv=\xunit$, $\vv={\bf 0}$
and $\phi=0$, therefore
\begin{equation}
 \label{current_0}
 I=I^0
  =I_r^0\cos(\omega t)-I_i^0\sin(\omega t)
  =\frac{\sqrt{2}V S}{d}\
   [\sigma_\perp\cos(\omega t)-\epsilon_\perp\omega \sin(\omega t)]\ .
\end{equation}
In the convecting state all fields $\nv, \vv$ and $\phi$ are modified,
as is $I$.
Within the SM, for homogeneous stationary roll solutions
\begin{equation}
 \label{current_2}
 I=I_r\cos(\omega t)-I_i\sin(\omega t)
  +\textit{higher temporal harmonics}
\end{equation}
where the amplitudes of the higher temporal harmonics are expected to be much
smaller than $I_r$ and $I_i$~,
at least at intermediate frequencies
$\omega\simeq 1/\tau_0$
where $\tau_0=\epsilon_\perp/\sigma_\perp$ is
the charge-diffusion time.
We define the real and imaginary reduced Nusselt numbers, $\Nur$ and $\Nui$~,
as $I_r/I_r^0-1$ and $I_i/I_i^0-1$~, respectively.
Thus $\Nur=\Nui=0$ in the quiescent state, while in the convecting state
$\Nur$ measures the excess
energy dissipation due to convection of the nematic liquid crystal, that is
the time average
$\left<\sqrt{2}V\cos(\omega t) I(t)\right>_t ~ =~(1+\Nur) V^2\sigma_\perp
S/d$.
In heuristic terms the effective resistance of the nematic layer
is changed by convection from $R_0=d/(\sigma_\perp S)$ to
$R=R_0/(1+\Nur)$ ;
equivalently the imaginary Nusselt number measures the change in the
effective capacitance of the nematic layer, $C_0=\epsilon_\perp S/d$
in the quiescent state, $C=C_0(1+\Nui)$ in the convecting state.
When the reduced distance from onset
$\epsilon~\equiv ~  V^2 / V_c^2-1$ is small,
the electric Nusselt numbers can be calculated for homogeneous rolls
using weakly nonlinear methods.
Assuming that the leading convection amplitude,
$A$, associated with the linear roll mode,
remains small,
a systematic expansion in powers of $A$ is performed.
After adiabatic elimination of the slave modes
and calculation of the resonant saturating cubic terms
approximate roll
solutions are obtained together with the relation
$A(\epsilon)=a_0\sqrt\epsilon$~.
The current can then be calculated from \eqref{current_gene}.
For symmetry reasons the first contribution from the convection modes
comes at order $A^2$, and therefore one expects
$\Nur\sim A^2\sim\epsilon~,\ \Nui\sim A^2\sim\epsilon$
in the weakly nonlinear regime.
That is, Nusselt numbers allow a direct measurement of the convection
amplitude $A$, and therefore a test of the supercritical law
$A(\epsilon)=a_0\sqrt\epsilon$~.
In order to make this clear and to obtain
\textit{approximate} analytic formulae,
one can first use the quasi-unidimensional approximation,
where all fields are considered at the middle of the layer ($z=0$)
and only their $x$-dependence is kept.
The linear normal roll mode then assumes the form
$n_z=-A N_z\sin(q x),
v_z=A/(q\tau_0) V_z \cos(q x),
\phi=A/(qd)[\Phi_c\cos(\omega t)+\Phi_s\sin(\omega t)]\cos(q x)$
where, as in the rest of our theoretical calculations,
we only keep the lowest nontrivial time-mode for each field.
We also choose as a normalization condition $N_z=1$~;
then $V_z~, \Phi_c$ and $\Phi_s$ are calculated at fixed frequency
by solving the linear neutral eigenproblem.
With $\rho_e=\nabv \cdot \Dv$, $j_z+\partial_t D_z$
can be easily calculated at $z=0$.
Keeping only the horizontally homogeneous terms because of the surface
integral in \eqref{current_gene}, one obtains to lowest order in $A$
\begin{eqnarray}
\label{Nur}
\Nur&=&\frac{A^2}{2}\Big[
\sigma_a' N_z (N_z-\Phi_c)+\epsilon_\parallel'\Phi_c V_z
-\epsilon_a' N_z (V_z+\omega \tau_0\Phi_s)\Big]\\
\label{Nui}
\Nui&=&\frac{A^2}{2}\Bigg[
\epsilon_a' N_z (N_z-\Phi_c)
+\frac{\Phi_s}{\omega \tau_0}\ (\sigma_a' N_z-\epsilon_\parallel' V_z)\Bigg]
\end{eqnarray}
with
$\sigma'_a=\sigma_a/\sigma_\perp~,\
\epsilon'_a=\epsilon_a/\epsilon_\perp$ and
$\epsilon'_\parallel=\epsilon_\parallel/\epsilon_\perp$~.
For standard nematic materials with large positive $\sigma'_a$
(see e.g. \eqref{elecparams}),
the leading term in $\Nur$ \eqref{Nur}
is the anisotropic conduction term in $\sigma'_a N_z^2$,
which imposes a positive value of $\Nur$~.
One thus expects that the
tilt of the director out of the plane in roll structures
will enhance the electrical conduction of the layer
and finally the in-phase current.
Concerning $\Nui$ \eqref{Nui}
it should be noted that $\Phi_s/\omega$ tends to a finite positive value
when $\omega\rightarrow 0$.
Two terms of \eqref{Nui} control the sign of $\Nui$~.
The first term
in $\epsilon'_a N_z^2$
reveals a diminution of the effective capacitance of the cell
due to the director tilt, since
the dielectric anisotropy $\epsilon'_a$ of
the nematic materials used in electroconvection
is usually negative.
The other important contribution
is the positive term in
$\sigma'_a N_z \Phi_s/\omega$~,
which expresses that the potential modulation induced by the convection
creates by coupling with the director tilt an out-of-phase current
$I_i>0$ (see \eqref{current_2}).
Since $\Phi_s/\omega$ decreases with $\omega$,
this positive term can compensate the negative term in $\epsilon'_a N_z^2$
at low frequencies only, and for nematics with large $\sigma_a'$~.
A \textit{numerical} calculation, 
using a standard Galerkin technique to expand
the $z$-dependence of all fields in test functions,
can provide a more accurate evaluation of $\Nur$ and $\Nui$~.
For this purpose we have modified the code developed in \cite{plaut-pesch-99}
to use Tchebyshev polynomials as the test functions in order to accelerate
the convergence (typically 4 $z$-modes were sufficient),
and inserted a procedure to calculate the Nusselt numbers.
For convenience the current
\eqref{current} is evaluated at the lower plate $z=-d/2$ where,
because of the boundary conditions,
$j_z+\partial_t D_z$ reduces to $(\sigma_\perp+\epsilon_\perp\partial_t)E_z$.
Thus, since the convection induced potential is even under
$z\mapsto-z$ at linear order, but odd at quadratic order,
one sees that
the leading contribution to $I$ comes from the potential part of the
homogeneous quadratic slave mode noted
$A^2 V_2(\qv,-\qv)$ in eq. (27) of
\cite{plaut-pesch-99}.
Of course the saturation at cubic order needs also to be calculated
in order to provide the law $A=a_0\sqrt\epsilon$.
We will return to the numerical results (Fig. \ref{comparison}),
which confirm the trends found from the analytic formulae
eqs. (\ref{Nur}), (\ref{Nui}),
after presenting our experimental results.

We use the ``classical'' arrangement \cite{Joets}
based on a pre-fabricated liquid crystal
cell\cite{EHC}.
The glass plates have
no spacers, glue, etc. within the active area;
they are separated by
$d ~=~ 23.4 \pm 0.5 \mu$m. 
With air only between the plates,
we measure, using a auto-balancing 1kHz bridge,
the capacitance of the
cell in order to determine accurately  (within 8 ppm)
the ratio $S/d$
(nominally $S=5 \mathrm{mm}\times 5 \mathrm{mm}$).
After this measurement the nematic liquid crystal
methoxy-benzylidene butyl-aniline (MBBA) \cite{Delta Tech}, doped with
0.0005\% tetrabutyl ammonium bromide, is
introduced between
the transparent conducting electrodes.
The filled cell is placed
in a temperature controlled housing,
and then introduced between the pole faces of a
large electromagnet. As the
nematic liquid crystal undergoes
the magnetically induced splay
Frederiks transition,
the capacitance and conductance of the cell are monitored.
>From these measurements we obtain
both electric conductivities and both dielectric constants
\cite{Freederickztransition}.
Hence, we measure {\em in situ}
all the electrical transport properties of the specific
nematic liquid crystal used.
For the experiments reported here, all at $28^\circ$C, we find
\begin{eqnarray}
\sigma_\perp=(8.5\pm 0.8)\ 10^{-8} (\Omega \mathrm{m})^{-1} &\; ,\; &
\sigma_a'=\sigma_a/\sigma_\perp=0.35\pm 0.04\; ,\nonumber\\
\label{elecparams}
\epsilon_\perp = (4.65\pm 0.03)\epsilon_0 &\; ,\; &
\epsilon_a'=\epsilon_a/{\epsilon_\perp} = -0.080\pm 0.001\; ,
\end{eqnarray}
with $\epsilon_0$ the vacuum dielectric permittivity.
After these measurements, the
nematic cell is transferred to the stage of
a polarizing microscope so that shadowgraph
\cite{shadowgraph}
images can be obtained concomitantly
with the electric current measurements.
A function generator is used to produce
a sinusoidal voltage signal which is in turn amplified, and applied
to the cell.
The path to ground for the current traversing the
cell is through a current-to-voltage converter.  The
output signal from this converter is measured by a lock-in amplifier, whose
reference signal is supplied by the original function generator.
Before any measurements are taken, the
nematic cell is replaced by a purely resistive load and
the phase setting on the lock-in is adjusted to zero the out-of-phase
current component. The
nematic cell is then re-inserted.
Then, at a selected frequency $V$ is raised in small steps.
At each step, after waiting several seconds, $I_r$ and $I_i$ are recorded
\cite{prec_curr_expe}.
This proceeds until a maximum desired $V$ (well above
the threshold value $V_c$) is reached.
Then, the process is reversed, and the currents recorded as
$V$ decreased. The difference in current for increasing vs decreasing
$V$ is less than 2\%.
When $V$ is raised above $V_c$~,
the electric current traversing the liquid crystal measurably deviates from
its value in the quiescent state, $I^0$.
In order to determine $V_c$ from either the in-phase or out-of-phase
current data, we first determine a baseline for
$I_r^0$ ($I_i^0$) by fitting a straight line to $I_r$ ($I_i$) vs $V$ for
$V$ much smaller than
$V_c$~; see Fig. \ref{current}.
These values of $I_r^0/V$ and $I_i^0/V$ provide
independent measurement of $\sigma_\perp$ and $\epsilon_\perp$
(see \eqref{current_0}) that agree within
5\% with the direct measurement of these parameters using the Frederiks
transition.
The Nusselt numbers as functions of $V$ are then calculated by subtracting
unity from the 
ratios $I_r/I_r^0$ and $I_i/I_i^0$~.
By fitting another straight line to
$\Nu_r$ ($\Nu_i$) in the region where
it deviates from zero,
we define $V_c$ as where this line crosses zero
(see the insets in Figs. \ref{current} and \ref{nusselt1}).
For one ramp of $V$,
the three values of $V_c$ determined from the Nusselt
numbers and the traditional shadowgraph technique agree with each other within
0.01\%.

Our apparatus did not reach
sufficiently large $V$ to measure the crossover
to the dielectric regime (see e.g. \cite{kramer1}); we therefore estimated a 
characteristic ``cutoff'' frequency $\omega_c$ by
fitting $V_c(\omega)$ to the function $A/(\omega - \omega_c)$.
We found typically
$\omega_c/(2\pi)=645 \mathrm{Hz}$,
but during the course
of taking the measurements (2-3 months) this quantity varied
by $\pm 10\%$, probably in connection with variations of the
electrical parameters, especially the conductivities.

In Fig. \ref{nusselt1} we plot both 
the real and imaginary Nusselt numbers vs $\epsilon$. 
While $\Nur$ is
always observed to be positive, for the data set shown $\Nui$ is negative.
In some cases
(discussed subsequently) it becomes positive.
Note also that $\Nur$ is at least
ten times larger in magnitude than $\Nui $.

Close to threshold, i.e. for $0\le\epsilon\lesssim 0.1$,
both Nusselt numbers are proportional to $\epsilon$
as shown in the inset for the real Nusselt number:
this confirms the supercritical law $A\sim\sqrt\epsilon$.
The variations of the corresponding slopes $\Nur/\epsilon$
and $\Nui/\epsilon$ vs $\omega\tau_0$ are given on Fig. \ref{comparison},
which represents the results of several ramps in $\epsilon$ at each frequency.
For $\epsilon\gtrsim 0.1$
the curves deviate from straight lines;  the ``knees''
in the curves indicate clearly
the onset of secondary instabilities.  Specifically,
the two arrows shown on Fig. \ref{nusselt1} 
correspond to the onset potential differences for
the zig-zag instability
\cite{RibJoeLin-ZZ-86}
and the spontaneous generation of dislocations
associated with the so-called
``defect chaos''
\cite{BrauRasStein-91}.

To compare our experimental results with the SM calculations,
we used the elastic constants and the viscosities
measured for MBBA at $28^\circ\mathrm{C}$ in \cite{MBBAelast,MBBAvisc},
and the electric parameters that we determined independently
(\eqref{elecparams}).
Varying those parameters within the stated uncertainties,
we calculate (with the fitting procedure defined above)
for the ``cutoff'' frequency $\omega_c/(2\pi)=730 \pm 120 \mathrm{Hz}$,
in agreement with the measured value.
Systematically varying the electrical parameters
within the experimental error bars,
we also calculate, with the weakly nonlinear numerical code introduced above,
the
bands of possible values of $\Nur/\epsilon$
and $\Nui/\epsilon$.
These bands are drawn in gray on Fig. \ref{comparison}.
The extremal values turn out to be obtained by variation of only $\sigma'_a$,
with the upper (lower) curves for both $\Nur/\epsilon$ and $\Nui/\epsilon$
obtained for the largest (smallest) value of $\sigma'_a$~.
This is consistent with the fact that the leading positive terms
controlling the Nusselt numbers
as seen in the approximate formulae eqs. (\ref{Nur}), (\ref{Nui})
are proportional to $\sigma'_a$~.
Note also that for large $\sigma'_a$ we expect $\Nui$ to be positive
at low frequencies, while $\Nui$ is always negative for $\sigma'_a$ small .
There is a good agreement between experiments and theory concerning
the imaginary Nusselt number $\Nu_i$~;
on the other hand the real Nusselt number $\Nur$ decreases more abruptly
in the experiments than in the theory.
However the agreement obtained for $\Nui$ at all frequencies and
$\Nur$ at small frequencies is particularly
significant since (contrarily to the standard approach in nematic
electroconvection where usually $\sigma_\perp$ is fitted)
no adjusted parameters have been used.

In conclusion, electric Nusselt number measurements are validated
as a new and powerful method of characterization
of electroconvection.
This wholly quantitative technique stands in contrast
to traditional optical methods which only become quantitative
in certain limit cases.
This technique affords a precise determination of the threshold voltage 
for the onset of electroconvection as well as secondary instabilities.
Nusselt number measurements also represent an important quantitative
tool for testing competing theoretical descriptions of electroconvection.
Here, with only the limitation of relying on tabulated
values of some material parameters,
we have shown that the standard model for electroconvection
gives satisfactory predictions of the Nusselt numbers near onset.
One conspicuous explanation for 
the remaining discrepancies may be 
that nematic liquid crystals are quite clearly electrolytic conductors,
and thus the Ohmic conduction assumed in the standard model
introduces an important approximation.
Thus, it clearly is of interest
to extend the calculations presented here
within the so-called weak-electrolyte model \cite{treiber}.
Future directions of this work include
also employing liquid
crystal materials for which the applicability of the weak
electrolyte model has been established,
and experiments in the highly nonlinear, dynamical
scattering regimes that occur at very large $\epsilon$ \cite{kai}.

\begin{center}
\large ACKNOWLEDGMENTS
\end{center}
J. T. G. and N. G.
acknowledge
technical assistance from A. R. Baldwin.
Their work was supported in part by
Kent State University and the Ohio Board of Regents.
E. P. thanks B. Dressel and W. Pesch for
very fruitful discussions and comparisons with their code.

\begin{figure}
\caption{
In-phase current $I_r$ vs the applied voltage $V$ at
a frequency $\omega/(2\pi) ~=~ 100\mathrm{Hz}$
i.e. $\omega\tau_0=0.30$.
{\em Inset:} blowup of the neighbourhood of $V_c$~.
}
\label{current}
\end{figure}

\begin{figure}
\caption{
Real and imaginary reduced Nusselt numbers vs the distance to threshold
$\epsilon$ for electroconvection of MBBA
at the same frequency than Fig. \ref{current}.
The arrow on the left indicates the onset of the
secondary zig-zag instability
and the arrow on the
right indicates the onset of
defect chaos.
{\em Inset:}
blowup of the $\epsilon \sim 0$ region indicating
how $\Nur$ initially increases linearly with $\epsilon$.
Small pre-transitional effects
caused by sample imperfections are also seen.
}
\label{nusselt1}
\end{figure}

\begin{figure}
\caption{
Black circles: measured
ratios $\Nur/\epsilon$ and $\Nui/\epsilon$
at small $\epsilon$
vs the dimensionless frequency. 
The vertical error bars correspond
to the imprecision in measuring $\Nu/\epsilon$
and to the dispersion between different $\epsilon$ ramps, 
while the horizontal error bars originate from the variation in
the charge-diffusion time
$\tau_0$~.
Gray bands: intervals of values of the same quantities deduced from the
weakly nonlinear SM calculations,
taking into account the variations of the electrical parameters of the
nematic liquid crystal
\eqref{elecparams}.
The upper (lower) theoretical curves
on both graphs were calculated with the largest (smallest)
value of $\sigma_a'=\sigma_a/\sigma_\perp$.
}
\label{comparison}
\end{figure}


\begin{references}
\bibitem{cross}
For a review on pattern formation, see
Cross M.C. and Hohenberg P.C., {\it Rev. Mod. Phys.},
{\bf 65} (1993) 851.

\bibitem{benard-rayleigh}
B\'enard H., {\it Annales de Chimie et de Physique}, {\bf 23} (1901) 62;
Rayleigh L., {\it Phil. Mag.}, {\bf 32} (1916) 529.

\bibitem{standardmodel}
Helfrich W., {\it J. Chem. Phys.}, {\bf 51} (1969) 4092;
Dubois-Violette E., de Gennes P. G. and Parodi O., {\it J. Phys. (France)},
{\bf 32} (1971) 305.


\bibitem{kramer1}
Kramer L. and Pesch W. in {\em Pattern Formation in Liquid Crystals},
Buka A. and Kramer L., eds,
(Springer, New York) 1996.

\bibitem{schmidt}
Schmidt R. J. and Milverton S. W., {\it Proc. Roy. Soc. (Lond.)},
{\bf A152} (1935) 486.

\bibitem{ahlers1}
Liu J. and Ahlers G., {\it Phys. Rev E}, {\bf 55} (1997) 6950.

\bibitem{chavanne}
Chavanne X. et al, {\it Phys. Rev. Lett.}, {\bf 79} (1997) 3648.

\bibitem{Rehberg1}
Rehberg I., Winkler B.L., de la Torre-Juarez M.,
Rasenat S. and Sch\"{o}pf W.,
{\it Festk\"{o}rperprobleme}, {\bf 29} (1989) 35.

\bibitem{plaut-pesch-99}
Plaut E. and Pesch W.,
{\it Phys. Rev. E}, {\bf 59} (1999) 1747.

\bibitem{treiber}
Dennin M., Treiber M., Kramer L., Ahlers G. and Cannell D.,
{\it Phys. Rev. Lett.}, {\bf 76} (1996) 319.


\bibitem{Joets}
A. Joets and R. Ribotta, {\it Phys. Rev. Lett.}, {\bf 60} (1988) 2164.

\bibitem{EHC}
E.H.C. Co, Tokyo, Japan.

\bibitem{Delta Tech}
Delta Technologies, Stillwater, MN.

\bibitem{Freederickztransition}
At zero magnetic field
$H$, the cell's resistance is $R=d/(\sigma_\perp S)$
and the capacitance $C=\epsilon_\perp S/d$;
as $H$ becomes larger than the Frederiks
threshold value, we fit the
$R(H), C(H)$ data to
obtain the values of $\sigma_a', \epsilon_a'$.

\bibitem{shadowgraph}
Rasenat S., Hartung G., Winkler B. L. and Rehberg I., {\it Exp.  Fluids},
{\bf 7} (1989) 412;
Joets A. and Ribotta R.,
{\it J. Phys. I France}, {\bf 4} (1994) 1013.

\bibitem{prec_curr_expe}
After this delay $I_r$ and $I_i$ always reach a stationary value.
We also use the digital signal processing capability 
of the lock-in amplifier to measure
higher temporal harmonics of the current;
these are found to always be at least two
orders of magnitude smaller than the fundamental.

\bibitem{RibJoeLin-ZZ-86}
Ribotta R., Joets A. and Lin Lei, {\it Phys. Rev. Lett.},
{\bf 56} (1986) 1595.

\bibitem{BrauRasStein-91}
Braun E. , Rasenat S. and Steinberg V.,
{\it Europhys. Lett.}, {\bf 15} (1991) 597.

\bibitem{MBBAelast}
de Jeu W. H., Claassen W. A. P. and Spruijt A. M.,
{\it Mol. Cryst. Liq. Cryst.},
{\bf 37} (1976) 269.
\bibitem{MBBAvisc}
Kneppe H., Schneider F. and Sharma N. K.,
{\it J. Chem. Phys.}, {\bf 77} (1982) 3203.

\bibitem{kai}
Kai S. and Hirakawa K., {\it Supp. Prog. Theor. Phys.}, {\bf 64} (1978) 212.

\end{references}
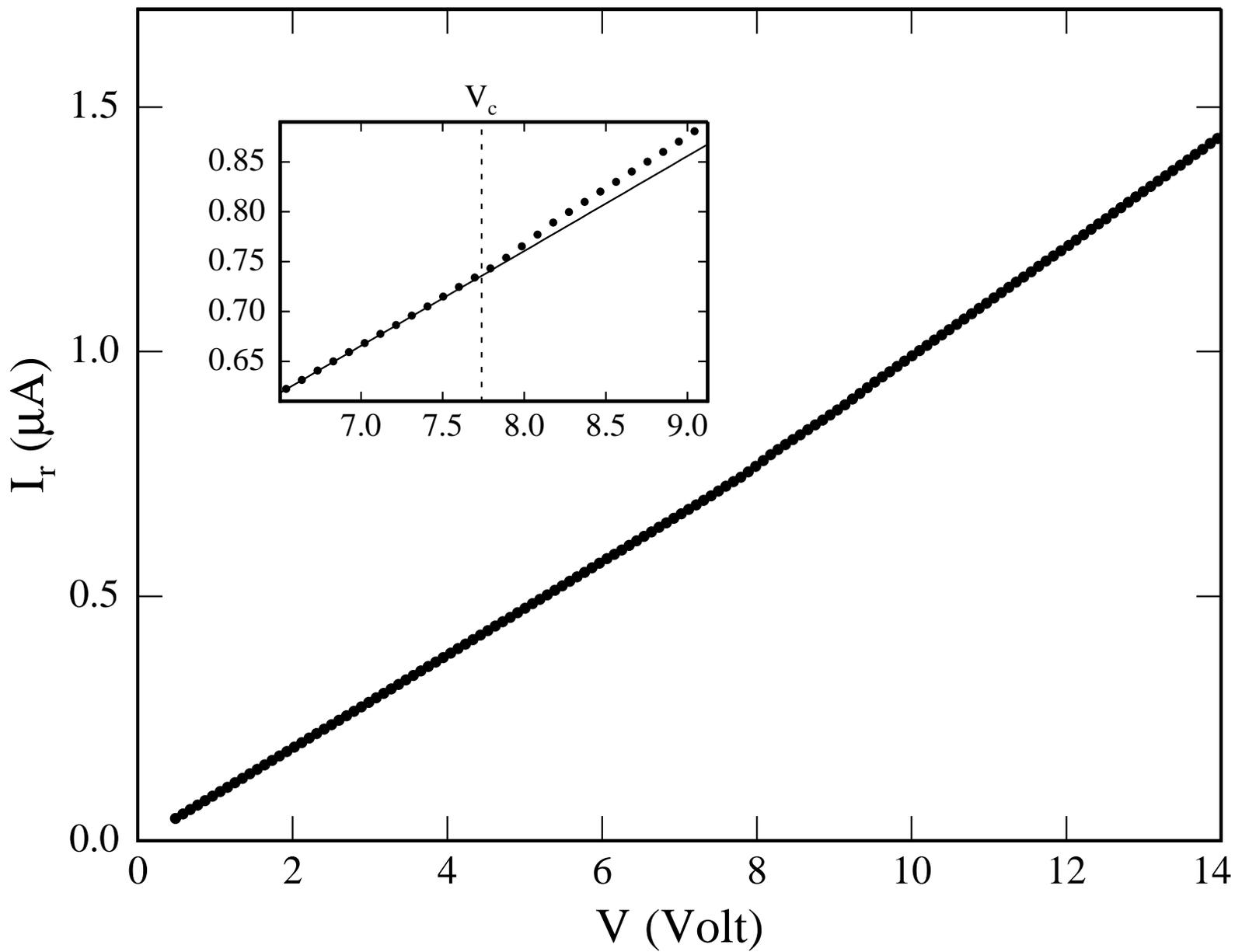
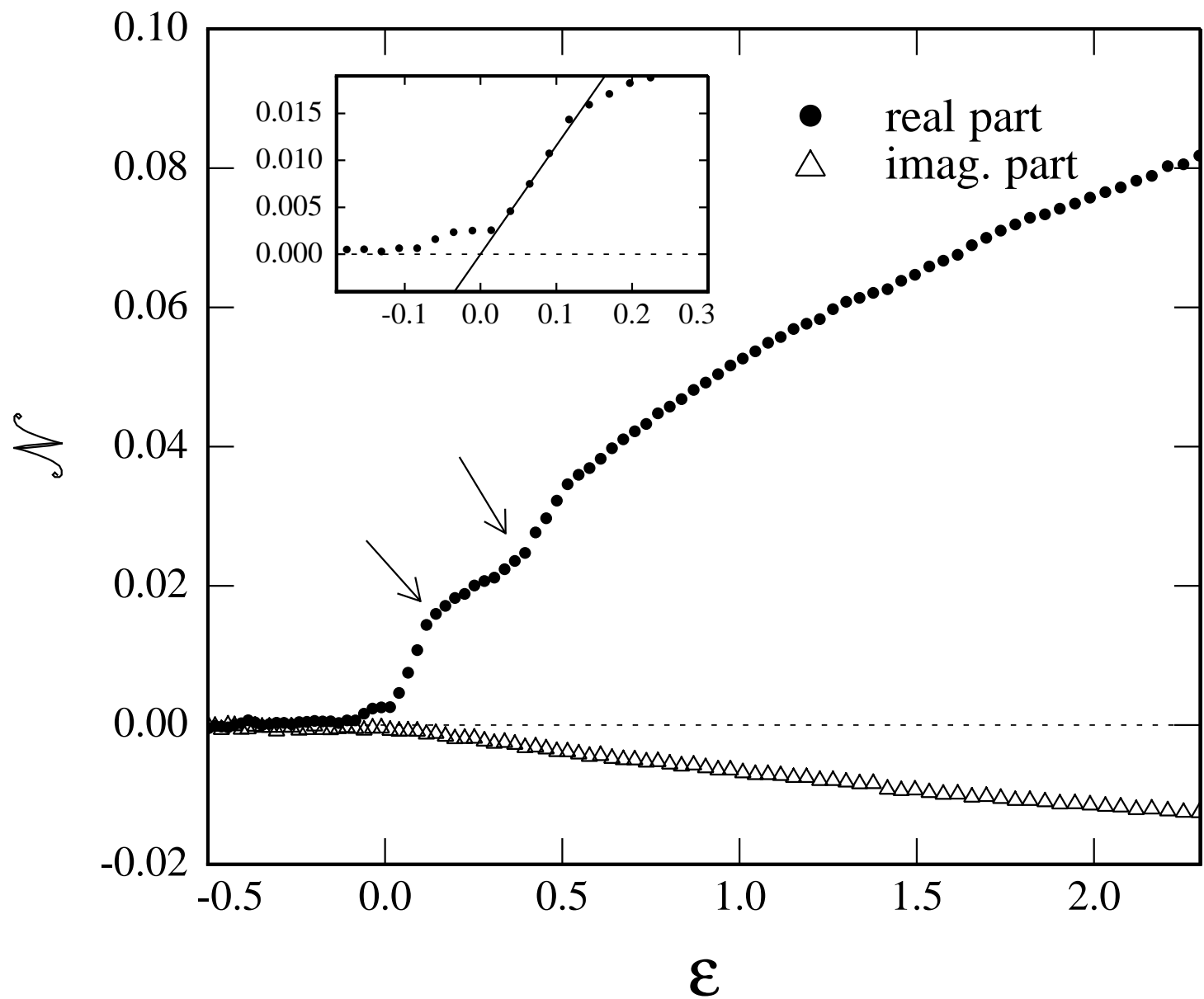
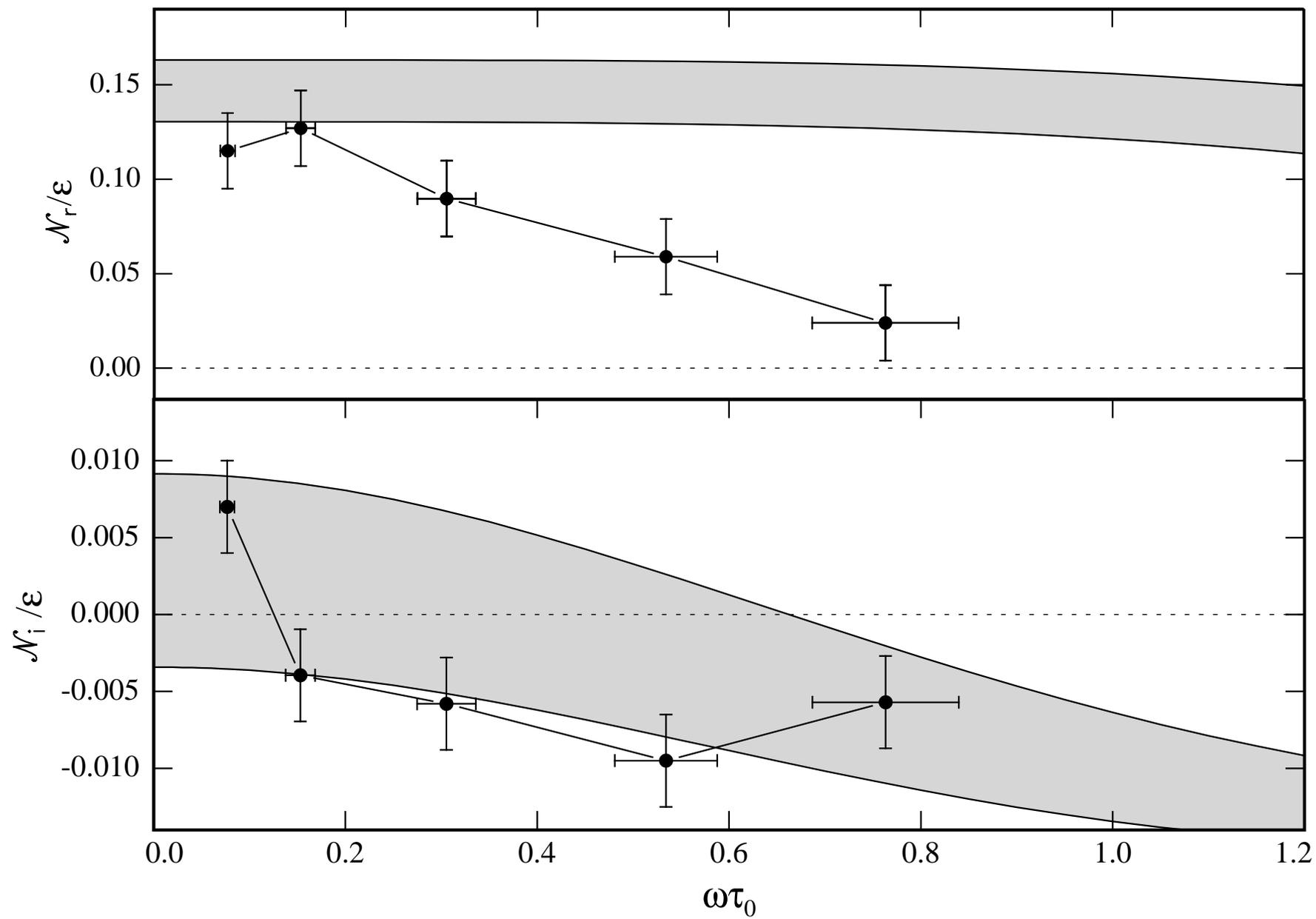
\end{document}